\documentclass[9pt,conference]{IEEEtran}
\IEEEoverridecommandlockouts
\usepackage{cite}
\usepackage{amsmath,graphicx}

\usepackage{multirow}
\usepackage{bbding}
\usepackage{booktabs}
\usepackage{cite}
\usepackage{verbatim}

\usepackage[utf8]{inputenc}
\usepackage{makecell}
\usepackage{tipx}

\newcommand{\xmark}{\ding{55}}

\usepackage{color}
\usepackage{hhline}
\usepackage{subfigure}
\usepackage{pifont}
\usepackage{bm}
\usepackage{amsfonts,amssymb,mathtools}

\usepackage{graphicx,float}
\usepackage[table]{xcolor}

\usepackage{amsmath,amssymb,amsfonts}
\usepackage{algorithmic}
\usepackage{graphicx}
\usepackage{textcomp}
\usepackage{xcolor}

\def\BibTeX{{\rm B\kern-.05em{\sc i\kern-.025em b}\kern-.08em
    T\kern-.1667em\lower.7ex\hbox{E}\kern-.125emX}}

\title{Structured Speaker-Deficiency Adaptation of Foundation Models for Dysarthric and Elderly Speech Recognition
}
\author{\IEEEauthorblockN{Shujie Hu\textsuperscript{1}, 
Xurong Xie\textsuperscript{2}, 
Mengzhe Geng\textsuperscript{3}, 
Jiajun Deng\textsuperscript{1}, 
Zengrui Jin\textsuperscript{1}, 
Tianzi Wang\textsuperscript{1}, 
Mingyu Cui\textsuperscript{1}\\ 
Guinan Li\textsuperscript{1}, 
Zhaoqing Li\textsuperscript{1}, 
Helen Meng\textsuperscript{1}, 
Xunying Liu\textsuperscript{1}}
\IEEEauthorblockA{\textit{$^{1}$The Chinese University of Hong Kong, Hong Kong SAR, China}\\
\textit{$^{2}$Institute of Software, Chinese Academy of Sciences, Beijing, China}\\
\textit{$^{3}$National Research Council of Canada, Canada}}}

\begin{document}
\maketitle

\begin{abstract}
Data-intensive fine-tuning of speech foundation models (SFMs) to scarce and diverse dysarthric and elderly speech leads to data bias and poor generalization to unseen speakers. This paper proposes novel structured speaker-deficiency adaptation approaches for SSL pre-trained SFMs on such data. 
Speaker and speech deficiency invariant SFMs were constructed in their 
supervised adaptive fine-tuning stage to reduce  undue bias to training data
speakers, and serves as a more neutral and robust starting point for test time unsupervised adaptation.
Speech variability attributed to speaker identity and speech impairment severity, or aging induced neurocognitive decline, are modelled using separate  adapters that can be combined together to model any seen or unseen speaker. 
%% and ``factored out'' 
%% during foundation model adaptive fine-tuning to multi-speaker in-domain data, \textcolor{red}{which produces a speaker-deficiency invariant model to serve as a \tim{neutral} and robust starting point for test time unsupervised adaptation. The sources of variability learned by adapters are 
%%``factored in'' and 
%% \tim{then} combined to model \tim{test speakers} during test time unsupervised adaptation}. 
Experiments on the UASpeech dysarthric and DementiaBank Pitt elderly speech corpora suggest structured speaker-deficiency adaptation of HuBERT and Wav2vec2-conformer models consistently outperforms baseline SFMs using either: a) no adapters; b) global adapters shared among all speakers; or c) single attribute adapters modelling speaker or deficiency labels alone by statistically significant WER reductions up to 3.01\% and 1.50\% absolute (10.86\% and 6.94\% relative) on the two tasks respectively. The lowest published WER of 19.45\% (49.34\% on very low intelligibility, 33.17\% on unseen words) is obtained on the UASpeech test set of 16 dysarthric speakers.
\end{abstract}


\vspace{-0.2cm}
\section{Introduction}
Despite the rapid progress of ASR technologies targeting normal and healthy users, their application to those suffering from speech disorders 
%% elderly and dysarthric speech 
remains a challenging task to date\cite{sehgal2015model,xiong2018deep,liu2021recent,ye2021development,geng2022speaker, yue2022acoustic, hu23b_interspeech, hu2022exploiting}.
%% Ageing presents enormous challenges to the health care worldwide. 
%Dysarthria is a common type of speech disorder caused by a wide spectrum of motor control conditions including cerebral palsy, amyotrophic lateral sclerosis, stroke and brain injuries. In a broader context, older adults who are experiencing neurocognitive decline, such as Alzheimer’s disease (AD), often exhibit clear patterns of speech impairment such as weakened neuro-motor control in speech production and imprecise articulation \cite{ konig2018fully, alzheimer20192019}. 
Neurocognitive disorders, such as Alzheimer’s disease (AD), are often found among older adults and manifest themselves in speech and language deficiency such as weakened neuro-motor control in speech production and imprecise articulation \cite{ konig2018fully, alzheimer20192019}.
ASR-based assistive technology plays a vital role in not only improving their quality of life and social inclusion, but also facilitating large-scale automatic early diagnosis of neurocognitive impairment and preventive care \cite{ferri2005global,rudzicz2014speech,zhou2016speech,mirheidari2019dementia,ye2021development}.%,toth2018speech}.
\par
Elderly and dysarthric speech bring challenges on all fronts to current deep learning based ASR technologies predominantly targeting non-aged, healthy adult users: \textbf{1) large mismatch} between such data and non-aged, healthy adult voices; \textbf{2) data scarcity} \cite{liu2021recent, 10584335}; and \textbf{3) data diversity} including accent or gender, and \textbf{speech deficiency} brought by: \textbf{1) speech disorders such as dysarthria}; and \textbf{2) aging induced neurocognitive decline} \cite{kodrasi2020spectro,smith1987temporal}. For example, dysarthric speakers of very low speech intelligibility exhibit more discriminative patterns of articulatory imprecision, decreased volume and clarity, changes in pitch, increased dysfluencies and slower speaking rate, while those diagonalized with mid or high speech intelligibility are closer to normal speakers.
Such diversity among dysarthric or elderly speakers hinders not only speaker-independent ASR system training or fine-tuning on such data, but also their fine-grained personalization to individual users’ voices.
%This leads to the difficulty in both speaker-independent ASR system training or domain fine-tuning on such data, and their fine-grained personalization to individual users’ voices.
This issue is even more challenging when fine-tuning self-supervised learning (SSL) speech foundation models (SFMs) \cite{baevski2020wav2vec, chen2022wavlm, hsu2021hubert} that contain a large number of parameters. 
\par

Recently, test time training \cite{gandelsman2022testtime, sun2024learning} has been successfully applied to large language models (LLMs) adaptation tasks. Such techniques are closely related to the test time unsupervised speaker adaptation techniques that were widely employed across generations of ASR systems \cite{anastasakos1996compact, gales1998maximum, swietojanski2016learning, 8682667} including both hybrid and end-to-end (E2E) models \cite{liu2021recent, geng23_interspeech}. However, their application to foundation models for ASR tasks remains under explored to date. 

Data-intensive supervised fine-tuning of speech foundation models to the highly scarce dysarthric and elderly speech training data rapidly leads to data bias and poor generalization to unseen speakers. In order to perform effective test time unsupervised adaptation of SSL models to arbitrary unseen speakers, it is vital to first construct a more speaker and speech deficiency invariant SSL model during their supervised fine-tuning stage to reduce the undue bias to training data speakers, and serves as a more neutral and robust starting point for test time unsupervised adaptation, akin to the use of well-established speaker adaptive training (SAT) \cite{anastasakos1996compact, gales1998maximum, 8682667, liu2021recent, deng2023confidence} approaches. Furthermore, SAT based supervised fine-tuning of SSL models using highly diverse dysarthric and elderly speech data of varying degrees of speech disorder severity and speaker level attributes, such gender and age \cite{liu2021recent, geng23_interspeech, geng2022speaker} requires their multifaceted sources of variability to be modelled using separate adapters in a structured manner.  
\par
To this end, novel structured speaker-deficiency adaptation approaches are proposed in this paper for SSL pre-trained speech foundation models. % Multiple sources of heterogeneity attributed to speaker identity and speech impairment severity (for dysarthric speech), or age and neurocognitive impairment severity (for elderly speech), are modelled using separately constructed adapters and 、、\textcolor{red}{represented}
% during foundation model adaptive fine-tuning to multi-speaker in-domain data. 
% This crucial design allows canonical models that are independent of speaker identity and \textcolor{red}{speech deficiency}, to be produced, and serve as the starting point for fine-grained test time unsupervised adaptation to user data. During test time unsupervised adaptation, the sources of variability learned by these adapters are combined to model any unseen speaker.
Speaker and speech deficiency invariant SFMs were constructed in their 
supervised adaptive fine-tuning stage (i.e., system training) to reduce undue bias to training data
speakers, and serves as a more neutral and robust starting point for test time unsupervised adaptation.
Speech variability attributed to speaker identity and speech impairment severity, or aging induced neurocognitive decline, are modelled using separate  adapters that can be combined together to model any seen or unseen speaker.
\par
Experiments on the benchmark UASpeech \cite{kim2008dysarthric} dysarthric and DementiaBank Pitt \cite{becker1994natural} elderly speech corpora suggest structured speaker-deficiency adaptation of large HuBERT \cite{hsu2021hubert} and Wav2vec2-conformer \cite{baevski2020wav2vec} models consistently outperforms baseline SFMs using either: a) no adapters; b) global adapters shared among all speakers; or c) single attribute adapters modelling speaker or deficiency labels alone by statistically significant WER reductions up to 3.01\% and 1.50\% absolute (10.86\% and 6.94\% relative) on the two tasks respectively.  
The lowest published WERs of 19.45\% (49.34\% on very low intelligibility, 33.17\% on unseen words) and 17.45\% are obtained after applying the proposed structured speaker-deficiency adaptation approach on a stronger baseline SFM \cite{hu2022exploring, 10584335} obtained using cross-system multi-pass rescoring.

The main contributions of the paper are summarized below:\\ 
\textbf{1)} To the best of our knowledge, this paper presents the first work to apply test time unsupervised adaptation to SFMs for dysarthric and elderly speech recognition. In contrast, previous researches utilizing test time unsupervised adaptation were conducted on hybrid and end-to-end (E2E) ASR systems \cite{liu2021recent, geng23_interspeech}, while speaker adaptation on SSL SFM was performed in a supervised manner \cite{baskar2022speaker}. \\
\textbf{2)} This paper presents the first work to apply supervised adaptive fine-tuning on SFMs to produce a more neutral and robust starting point for test time unsupervised adaptation. In contrast, previous studies on this approach were conducted on hybrid and E2E ASR systems \cite{liu2021recent, geng2022speaker, geng23_interspeech}. \\
\textbf{3)} This paper pioneers novel structured speaker-deficiency adaptation approaches for SFMs. In contrast, prior researches in this direction significantly differ from this work by either: \textbf{a)} using speaker identity alone in in-domain data trained non-SSL based ASR systems \cite{9053725, shor19_interspeech, liu2021recent, geng23_interspeech, geng2022speaker, qi2023parameter}, or SSL foundation model adaptation \cite{baskar2022speaker}; or \textbf{b)} using speaker-deficiency information when adapting non-SSL, traditional hybrid TDNNs \cite{geng23b_interspeech}. \\
\textbf{4)} The best performing structured speaker-deficiency adapted HuBERT produces the lowest published WER of 19.45\% (49.34\% on very low intelligibility, 33.17\% on unseen words) on the benchmark UASpeech test set.

\vspace{-0.1cm}
\section{SSL Pre-trained ASR Models}

Speech SSL models such as Wav2vec2.0 \cite{baevski2020wav2vec}, HuBERT \cite{hsu2021hubert}, and WavLM \cite{chen2022wavlm} share similar Transformer backbones with supervised models. For example, Wav2vec2.0 consists of three components, including 1) a multi-layer CNN-based feature encoder; 2) an $L$-layers transformer-based context network; and 3) a quantization module. In this paper, we fine-tune the pre-trained Wav2vec2.0 model and HuBERT model with a
pure CTC decoder.

% which encodes raw speech audio input $\bm{X}$ into continuous speech representations $\boldsymbol{z}_t \in \boldsymbol{Z}$ with a stride of 20 ms and a receptive field of 25 ms; 2) an $L$-layers transformer-based context network producing contextual representations $\boldsymbol{c}_t \in \boldsymbol{C}$ over a sequence of randomly masked feature encoder outputs; and 3) a quantization module generating discrete speech units%$\boldsymbol{q}_t \in \boldsymbol{Q}$ as \textbf{pseudo-labels} for SSL pre-training.
% \par
% \noindent
% \textbf{HuBERT} \cite{hsu2021hubert} pre-training alternates between two steps: 1) a clustering step to create pseudo-labels; and 2) a prediction step to produce labels for masked positions. The model architecture of HuBERT is similar to Wav2vec2.0, including a feature encoder, a $k$-means quantization module and a transformer-based context network followed by a projection layer.

\begin{figure}
    \centering
    \includegraphics[width=0.46\textwidth]{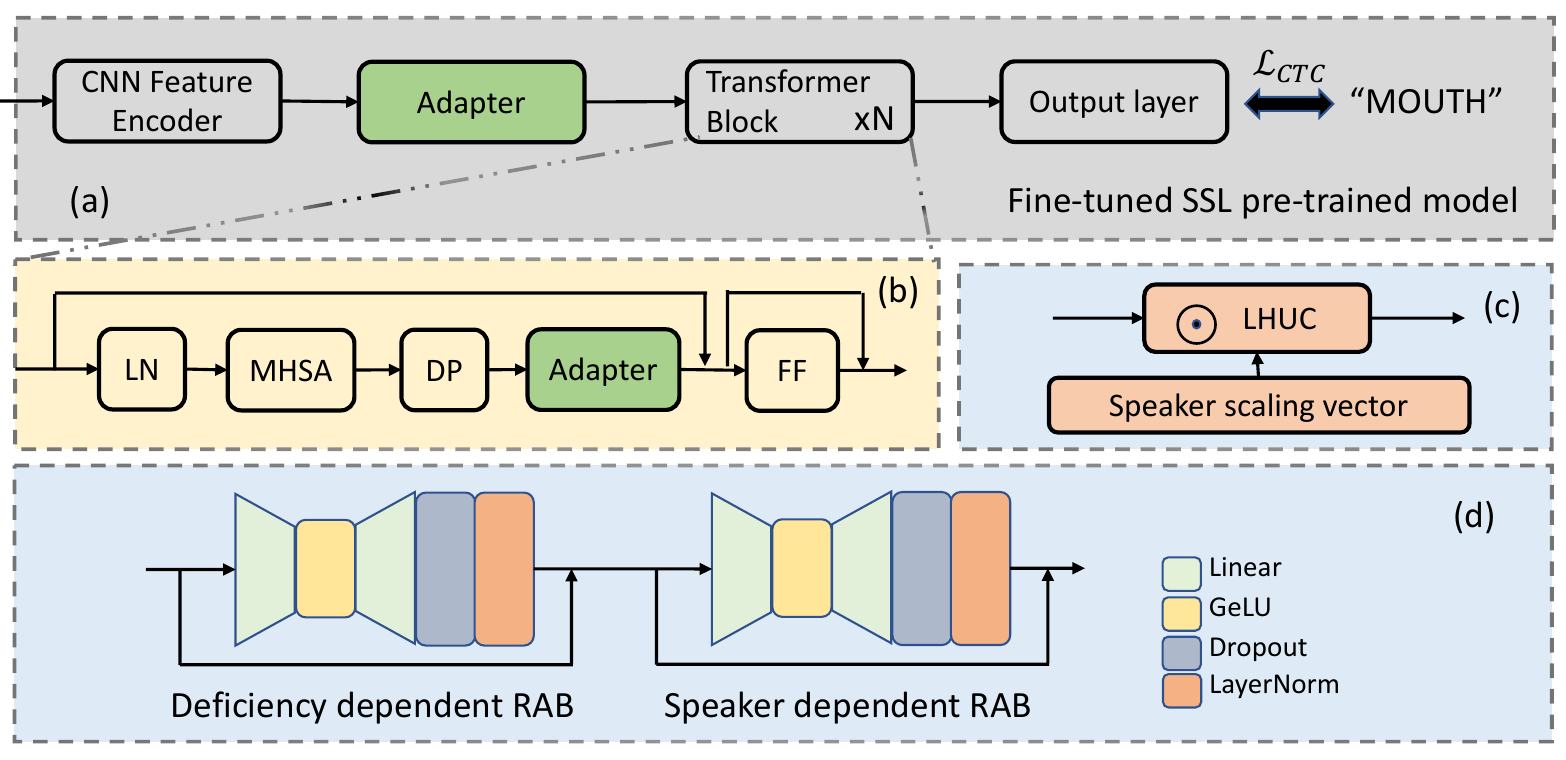}
    \vspace{-0.3cm}
    \caption{Examples of SSL pre-trained SFM (grey box, sub-figure (a)) adaptation using either speaker identity alone via \textbf{c)} speaker-dependent LHUC, or \textbf{d)} structured speaker-deficiency adapters. % separately modelling speech impairment severity for dysarthric speech, or age plus neurocognitive impairment severity for elderly speech, and also speaker identity. 
    The adapter can be inserted either \textbf{1)} after the CNN encoder (sub-figure (a)); or \textbf{2)} in a specific transformer block (sub-figure (b)).``LN'', ``MHSA'', ``DP'' and ``FF'' are layernorm, multi-head self-attention, dropout and feedforward modules.}.
    \label{fig:enter-label}
    \vspace{-0.5cm}
\end{figure}

\section{Adapter based Test Time Unsupervised Adaptation}
% 1.结构
% 2.可以表示整个测试集，。。。
\par
\noindent
\textbf{1. Adapter Architecture}:
\textbf{LHUC/HUB adapters} provide parameter-efficient and compact representations of the variability among dysarthric or elderly speakers \cite{liu2021recent, geng23_interspeech, swietojanski2016learning, abdel2013fast}. The key idea of LHUC adaptation is to use a speaker-dependent (SD) scaling vector to modify the amplitudes of activation outputs. 
Let $\bm r^{l,s}$ denotes the SD parameters for speaker $s$ in the $l$-th hidden layer, the hidden outputs adapted to the speaker $s$ is given as $\bm{h}^{l,s}=\xi(\bm{r}^{l,s}) \odot \bm{h}^{l}$.
% \vspace{-0.1cm}
% \begin{equation}
%     \label{eq:standard_lhuc}
%     \bm{h}^{l,s}=\xi(\bm{r}^{l,s}) \odot \bm{h}^{l}
%     \vspace{-0.1cm}
% \end{equation}
where $\bm{h}^{l}\in\mathbb{R}^{m}$ is the hidden outputs of $l$-th layer, $\odot$ denotes the Hadamard product operation, and $\xi(\cdot)$ is the element-wise $2 \times Sigmoid(\cdot)$ function. An example of incorporating LHUC into SSL pre-trained SFMs is shown in Fig. 1(c). HUB adaptation is similar to LHUC. Speaker-adapted hidden outputs can be given by $\bm{h}^{l,s}=\bm{r}^{l,s} + \bm{h}^{l}$.
%     \vspace{-0.2cm}
% \begin{equation}
%     \label{eq:standard_hub}
%     \bm{h}^{l,s}=\bm{r}^{l,s} + \bm{h}^{l}
%     \vspace{-0.1cm}
% \end{equation}
\par
\noindent
\textbf{Residual Adapter Blocks (RAB)} have been developed as a general parameter-efficient technique for pre-trained model fine-tuning \cite{thomas2022efficient, fan2022towards, 10023274, baskar2022speaker}. Let $f(\cdot;{\bm \Theta}_{l,s})$ denote the residual adapter function for speaker $s$ in the $l$-th layer, the adapted hidden outputs conditioned on the speaker are expressed by
\vspace{-0.1cm}
\begin{align}
    \label{eq:standard_adapter}
    \bm{h}^{l,s} &= f(\bm{h}^{l};{\bm \Theta}_{l,s}) + \bm{h}^{l}, \\
    f(\bm{h}^{l};{\bm \Theta}_{l,s}) & = \text{LN}(\text{DP}({\bm P}^{u}_{l,s}{\zeta({{\bm P}^{d}_{l,s} \bm{h}^{l}})}))
    \vspace{-0.1cm}
\end{align}
\noindent
where ${\bm \Theta}_{l,s}$ is the learnable SD parameters in the residual adapter. An RAB consists of a down-linear projection ${\bm P}^{d}_{l,s} \in \mathbb{R}^{k\times m}$, a GeLU activation function $\zeta(\cdot)$, an up-linear projection ${\bm P}^{u}_{l,s} \in \mathbb{R}^{m\times k}$, a dropout operation $\text{DP}(\cdot)$ and a layer-norm operation $\text{LN}(\cdot)$. 
% Examples of SSL pre-trained ASR model adapter fine-tuning using dysarthric or elderly speaker identity alone via either speaker-dependent LHUC adapters or RAB are shown in Fig. 1(a) and Fig. 1(b) respectively.
%Without loss of generality, the speaker identity encoded using LHUC and RAB can also be replaced by speech impairment severity, or age plus neurocognitive impairment severity labels for SSL model adaptation to speech deficiency alone. 
% An example of a RAB is shown in Figure 1(b).
% Compact LHUC, HUB, and RAB can be further used for speech deficiency adaptation by learning a speech deficiency-dependent (SDD) LHUC or HUB transform, or updating a specific adapter block. 

\par
\noindent
\textbf{Structured speaker-deficiency RAB} can separately model the rich sources of heterogeneity attributed to speaker identity and speech deficiency labels. % are modelled using separately constructed adapters. 
% and factored out during foundation model \textcolor{red}{adaptive training}. This produces speaker identity and speech deficiency independent, canonical models to serve as the starting point for fine-grained test time unsupervised adaptation to any seen or unseen speaker’s data that has no ground truth speech transcription or deficiency labels. 
% Taking factorised speaker-deficiency RAB adaptation of Fig. 1(d) as an example, 
As shown in Fig. 1(d), the final adapted hidden outputs are given by 
%In the cascaded factorised adaptation, the SDD parameters followed by SD parameters are cascaded into SSL pre-trained models. For example, without loss of generality, the cascaded RAB applied at the same layer are shown in Figure 1(c) and the factorised adapted hidden outputs can be given by
\vspace{-0.1cm}
\begin{equation}
    \bm{h}^{l,sd} = f(\bm{h}^{l};{\bm \Theta}_{l,sd}) + \bm{h}^{l},   \bm{h}^{l,sd,s} = f(\bm{h}^{l,sd};{\bm \Theta}_{l,s}) + \bm{h}^{l,sd}
\end{equation}
% \label{eq:factorised_adapter}
%     \bm{h}^{l,sd} &= f(\bm{h}^{l};{\bm \Theta}_{l,sd}) + \bm{h}^{l}, \\
%     \bm{h}^{l,sd,s} &= f(\bm{h}^{l,sd};{\bm \Theta}_{l,s}) + \bm{h}^{l,sd}, 
% \vspace{-0.2cm}
% \end{align}
where ${\bm \Theta}_{l,sd}$ denotes the speech deficiency conditioned adapter parameters for speakers labelled with a particular speech deficiency ``sd'' in the $l$-th hidden layer.

\par
\noindent
\textbf{2. Adapter Labels:} To investigate adaptation labels at different fine-grained levels, several adapter labels are used during test time unsupervised adaptation, including \textbf{a)} global level label, i.e., all speakers are classified into one category; \textbf{b)} speech impairment severity, or aging induced neurocognitive decline (speech deficiency) labels; \textbf{c)} speaker labels; and \textbf{d)} structured speaker-deficiency labels.
\par
\noindent
\textbf{3. Adapter Supervision and Estimation}: The overall procedure of test time unsupervised adaptation of foundation models using structured speaker-deficiency adapters is shown in Fig. 2(iii).
Let $\bm{\mathcal{D}}^{sd,s} = \{\bm{X}^{sd,s}, \bm{Y}^{sd,s}\}$ denote the data set for speaker $s$ whose speech deficiency label is $sd$, where $\bm{X}^{sd,s}$ and $\bm{Y}^{sd,s}$ are the waveform and the corresponding supervision token sequences, respectively.
Test time unsupervised adaptation is performed to speaker data initially without any speech transcription or speech deficiency labels provided. 
The speech deficiency labels of test data speakers are automatically predicted using the well-trained spectro-temporal embedding features based neural network classifiers \cite{geng23b_interspeech, geng2022speaker}.
The hypothesis supervision $\bm{\hat Y}^{sd,s}$ for adaptation is generated by initially decoding the test data using the baseline SSL foundation model fine-tuned to all the in-domain training data (Fig. 2(ii)).
The parameters of speech deficiency conditioned adapter  $\bm{\hat{\Theta}}_{sd}$, and speaker identity dependent adapters, $\bm{\hat{\Theta}}_{s}$, are estimated in turn using the CTC loss,

\begin{equation}
 \vspace{-0.2cm}
    \small
    \{\bm{\hat{\Theta}}_{sd}\} = \underset{\{\bm{\Theta}_{sd}\}}{ \arg\min} \{\mathcal{L}_{CTC}(\bm{\mathcal{D}}^{sd}; \bm{\Theta}_{sd})\} 
\end{equation}
\begin{equation}
    \small
    \{\bm{\hat{\Theta}}_{s}\} = \underset{\{\bm{\Theta}_{s}\}}{ \arg\min} \{\mathcal{L}_{CTC}(\bm{\mathcal{D}}^{sd,s}; \bm{\hat{\Theta}}_{sd}, \bm{\Theta}_{s})\} 
\end{equation}
where $\bm{\mathcal{D}}^{sd}$ is the union of all the dysarthric or elderly speakers' data that are labelled with a speech deficiency level of ``$sd$'', and $\bm{\mathcal{D}}^{sd,s}$ is the $s$ speaker's data labelled with a speech deficiency level of ``$sd$''.
%Test time unsupervised adaptation is performed to speaker data initially without any speech transcription or speech deficiency labels provided. 
%The speech deficiency labels of test data speakers are automatically predicted using the well-trained spectro-temporal embedding features based neural network classifiers \cite{geng23b_interspeech, geng2022speaker}. The hypothesis supervision $\bm{Y}^{sd,s}$ for adaptation is generated by initially decoding the test data using the baseline SSL foundation model fine-tuned to all the in-domain training data.
\begin{figure}
    \centering
    \includegraphics[width=0.43\textwidth]{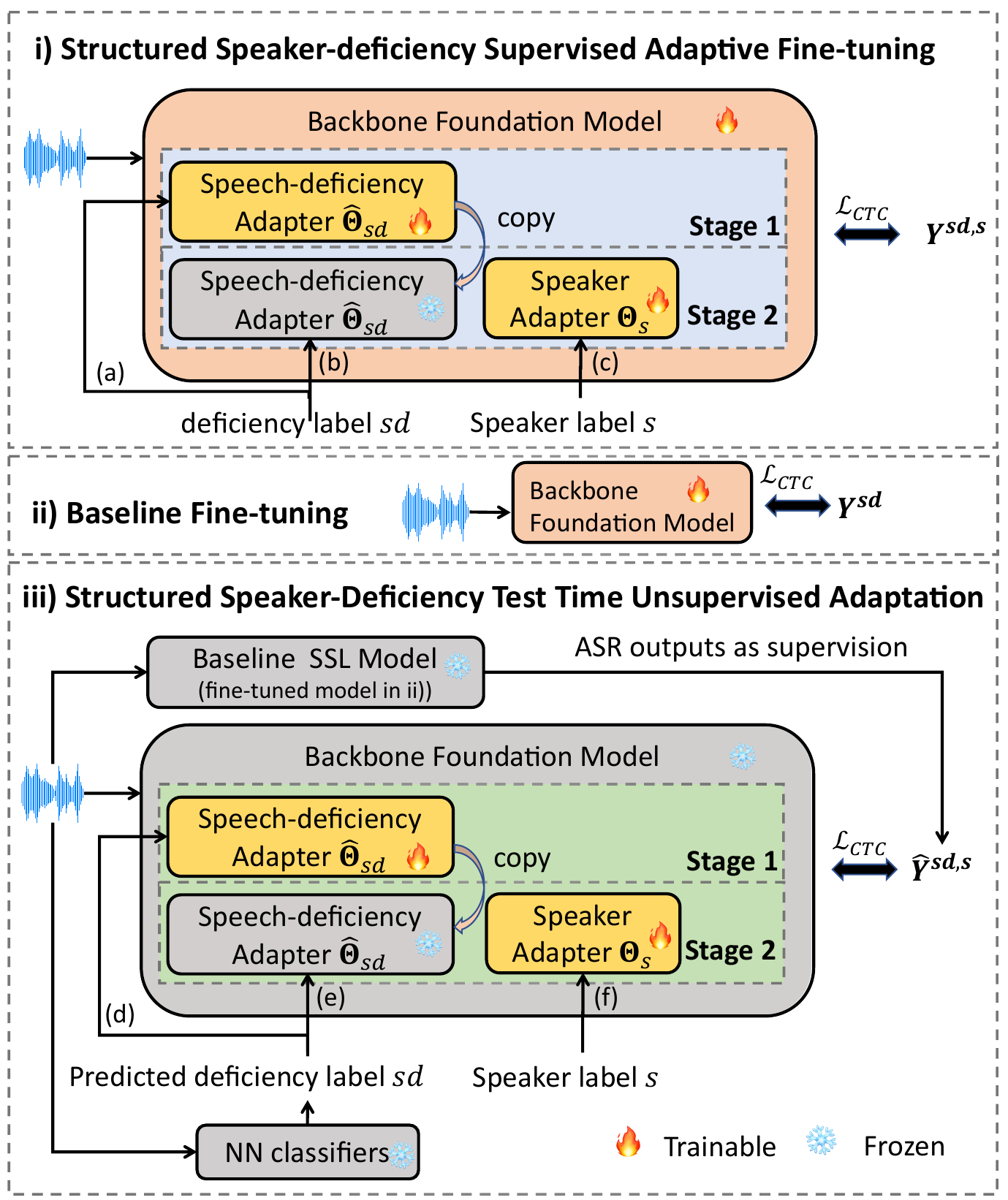}
    \vspace{-0.2cm}
    \caption{Examples of  \textbf{i)} adaptive training using structured speaker-deficiency adapters; \textbf{ii)} baseline fine-tuning; and \textbf{iii)} test time unsupervised adaptation using structured speaker-deficiency adapters. During adaptive training and test time adaptation, the parameters of the speech deficiency conditioned adapter and the speaker identity dependent adapter are estimated in turn in two stages.}
    \label{fig:stage}
    \vspace{-0.5cm}
\end{figure}

\section{Structured SFM Adaptation}

\subsection{Structured SFM Supervised Adaptive Fine-tuning}
Adaptive training (AT) \cite{anastasakos1996compact, gales1998maximum} is a well-established family of techniques that model heterogeneity in natural speech.
By representing factors of speech variability, e.g. speaker identity, using separate parameters based on, e.g. MLLR, CMLLR, LHUC transforms or adapters, AT produces canonical models independent of both speaker identity and speech deficiency during the supervised training stage, which serves as a more neutral and robust starting point than standard non-AT models for test time unsupervised adaptation. Their effectiveness has been widely demonstrated across generations of ASR systems from HMM, hybrid DNN to end-to-end systems \cite{anastasakos1996compact, gales1998maximum, swietojanski2016learning, deng2023confidence}. In our paper, the proposed form of structured speaker-deficiency supervised \textbf{adaptive fine-tuning (AFT)}, akin to the use of well-established adaptive training, enhances the speaker and speech deficiency invariance of SFMs.
% Adaptive training produces speaker identity and speech deficiency independent, canonical foundation models, which serve as a stronger starting point than standard non-AT models for test-time unsupervised adaptation.

The \textbf{adapter architectures} used during structured SFM supervised adaptive fine-tuning are also LHUC/HUB, RAB and structured speaker-deficiency RAB, while the \textbf{adapter labels} encompass speech deficiency labels, speaker level labels, and structured speaker-deficiency labels. The overall procedure of structured SFM supervised adaptive fine-tuning using speaker-deficiency adapters is shown in Fig. 2 (i). The \textbf{supervision} of supervised adaptive fine-tuning are ground truth speech transcript, and the speech deficiency and speaker identity conditioned adapters are optimized in turn and tightly integrated with the backbone foundation model parameters fine-tuned to multi-speaker in-domain training data.

\subsection{Test Time Unsupervised Adaptation on AFT Model} 
% The \textbf{adapter architectures}, \textbf{adapter labels} and \textbf{adapter supervision} are the same as those described in Sec. III. The foundation model after adaptive training is canonical, speaker identity and speech deficiency independent, which serves as a stronger starting point than standard non-AT models.
The \textbf{adapter architectures}, \textbf{adapter labels}, and \textbf{adapter supervision} are identical to those described in Sec. III. After structured supervised AFT, the canonical foundation model is independent of both speaker identity and speech deficiency, providing a more neutral and robust starting point compared to standard non-AFT models.

\begin{table}[h]
    \setlength\tabcolsep{2.2pt}
    \centering
    \caption{Performance of HuBERT systems without or with structured speaker (spk.) deficiency (defi.) adaptation
    %: \textbf{1)} baseline fine-tuning with no adaptation (Sys.1); \textbf{2)} non-structured speaker (spk.) only adaptation (Sys.2-10); \textbf{3)} non-structured deficiency (defi.) only adaptation (Sys.11-13); AND \textbf{4)} structured speaker-deficiency adaptation (Sys.14-17) 
    on \textbf{UASpeech}. LHUC or RAB are inserted at different layer positions (Pos.), where ``0'' stands for being after the CNN encoder (as shown in Fig 1(a)), while other numbers denote a position in the $x$-th transformer block (as shown in Fig 1(b)). ``VL/L/M/H'' refers to intelligibility subgroups. ``Param.'' is the number of adapter parameters. %where ``$a$M+$b$M'' denotes the number of speaker and speech deficiency adapter parameters being $a$ and $b$ millions, respectively. 
    ``Stru.'', ``Ada. Arc.'' and ``Sup.'' stand for ``Structured'', ``Adapter Architecture'' and ``Supervision''.} %``$\rightarrow$'' denotes multi-pass decoding \cite{hu2022exploring}. Sys. 0 is the TDNN joint decoding system using pre-trained HuBERT features  \cite{hu2022exploring}. $\dag$, $^\ast$ and $^\triangle$ denote statistically significant (MAPSSWE \cite{gillick1989some},$\alpha$ = 0.05) differences obtained against the baseline fine-tuned, non-structured speaker or deficiency alone adapted HuBERT systems (Sys. 1,9,13).}
    \vspace{-0.3cm}
    \scalebox{0.7}{
\begin{tabular}{c|c|c|c|c|cc|cccc|c|c} 
\hline\hline
\multirow{2}{*}{sys.} & \multicolumn{4}{c|}{Adaptive Training \& Test Time Adapt.}                                                                                                                        & \multicolumn{7}{c|}{WER (\%)}                                                                                                                                                                                                                                                                                                                                                                                                                                                                                                                                                         & \multirow{2}{*}{\begin{tabular}[c]{@{}c@{}}Ada.\\Param.\\(spk.+defi.)\end{tabular}}    \\ 
\cline{2-12}
                      & \begin{tabular}[c]{@{}c@{}}Ada.\\Arc.\end{tabular} & \begin{tabular}[c]{@{}c@{}}Ada.\\Pos.\end{tabular} & \begin{tabular}[c]{@{}c@{}}Ada.\\Label\end{tabular}            & Sup.       & unseen                                                                        & seen                                                                          & VL                                                                            & L                                                                             & M                                                                             & H                                                                            & O.A.                                                                                   &                                                                          \\ 
\hline\hline
1                     & \xmark                                                               & \xmark                                                               & \xmark            & \multirow{17}{*}{\begin{tabular}[c]{@{}c@{}}GT\\(for\\adaptive\\training);\\ \\ASR\\outputs\\of Sys. 1\\(for\\test time\\adaptation)\end{tabular}}                                                    & 50.06  & 13.30 & 59.47 & 33.62 & 22.22 & 6.34 & 27.71 & \xmark                                                                      \\ 
\cline{1-4}\cline{6-13}
2                     & \multirow{5}{*}{LHUC}                                               & \textbf{0}                                                                   & \multirow{9}{*}{spk.}                                    &          & \cellcolor{blue!25} 47.91                   & \cellcolor{blue!25} 13.31                              & \cellcolor{blue!25}60.02                          & \cellcolor{blue!25} 31.36                               & \cellcolor{blue!25} 20.98          & \cellcolor{blue!25} 5.96          & \cellcolor{blue!25} \textbf{26.88}     & \multirow{6}{*}{0.016M}                                                \\
3                     &                                                                     & 2                                                                   &                           &       & 48.61                   & 13.29                              & 59.89                          & 32.17                               & 21.43          & 5.93          & 27.14             &                                                                        \\
4                     &                                                                     & 7                                                                   &                              &     & 49.23                   & 13.41                              & 60.18                          & 32.56                              & 22.00                 & 6.07                 & 27.46             &                                                                        \\
5                     &                                                                     & 12                                                                  &                         &       & 48.90                   & 13.29                              & 59.61                          & 32.40                               & 21.71                 & 6.12                 & 27.25             &                                                                        \\
6                     &                                                                     & 18                                                                  &   &  & 48.88                   & 13.64                              & 60.32                          & 32.69                               & 21.51                 & 6.17                 & 27.46         &                                                                        \\ 
\cline{1-3}\cline{6-12}
7                     & HUB                                                                 & 0                                                                   &   &    & 48.78                   & 13.10                              & 59.25                          & 32.52                               & 21.22          & 6.04                 & 27.09         &                                                                        \\ 
\cline{1-3}\cline{6-13}
8                     & \multirow{3}{*}{RAB}                                                & 0                                                                   &   &    & 45.43                   & 13.16                              & 58.97                          & 29.98                               & 17.73 & 6.40                 & 25.81       & \multirow{3}{*}{8M}                                                  \\
9                     &                                                                     & \textbf{2}                                                                   &                                                              &         & \cellcolor{blue!25} 44.75          & \cellcolor{blue!25} 12.46             & \cellcolor{blue!25} 57.90         & \cellcolor{blue!25} 29.02                      & \cellcolor{blue!25} 18.24          & \cellcolor{blue!25} 5.46 & \cellcolor{blue!25} \textbf{25.12}    &                                                                        \\
10                    &                                                                     & 12                                                                  &   &    & 46.88                   & 12.80                       & 58.11                   & 31.45                               & 19.63          & 5.75          & 26.17             &                                                                        \\ 
\cline{1-4}\cline{6-13}
11                    & LHUC                                                                & 0                                                                   & \multirow{3}{*}{defi.}                                           &      & 48.35                   & 13.29                              & 59.63                          & 31.94                               & 21.41          & 5.99          & 27.04               & 4K                                                                  \\ 
\cline{1-3}\cline{6-13}
12                    & \multirow{2}{*}{RAB}                                                & 0                                                                   &   &    & 46.14                   & 13.06                              & 59.66                          & 30.21                               & 19.25         & 5.56          & 26.04           & \multirow{2}{*}{2M}                                                \\
13                    &                                                                     & \textbf{2}                                                                   &                                                               &        & \cellcolor{blue!25} 45.41          & \cellcolor{blue!25} 12.66             & \cellcolor{blue!25} 59.27                 & \cellcolor{blue!25} 29.79 & \cellcolor{blue!25} 17.94 & \cellcolor{blue!25} 5.30 & \cellcolor{blue!25} \textbf{25.51}        &                                                                        \\ 
\cline{1-4}\cline{6-13}
14                    & \multirow{4}{*}{\begin{tabular}[c]{@{}c@{}}Stru.\\RAB\end{tabular}}                                          & \textbf{0,0}                                                                 & \multirow{4}{*}{\begin{tabular}[c]{@{}c@{}}spk.+\\defi.\end{tabular}}                            &            & \cellcolor{blue!25} 44.60          & \cellcolor{blue!25} 11.87 & \cellcolor{blue!25} 57.38          & \cellcolor{blue!25} 28.41                  & \cellcolor{blue!25} 17.45      & \cellcolor{blue!25} 5.45          &  \cellcolor{blue!25} \textbf{24.70}        & \multirow{4}{*}{\begin{tabular}[c]{@{}c@{}}10M \\(8M+2M)\end{tabular}}                                         \\
15                    &                                                                     & 0,2                                                            &     &        & 45.06                   & 12.22              & 57.74          & 29.35                               & 18.06          & 5.32          & 25.10            &                                                                        \\
16                    &                                                                     & 2,0                                                            &     &         & 44.52 & 11.97          & 57.34 & 28.06         & 17.82          & 5.61        & 24.73        &                                                                        \\
17                    &                                                                     & 2,2                                                             &    &     & 45.11                  & 12.24              & 58.27          & 28.67                      & 18.59         & 5.30 & 25.13         &                                                                        \\
\hline\hline
\end{tabular}
}
    \label{tab:my_label}
    \vspace{-0.5cm}
\end{table}

\section{Experiments}
\vspace{-0.1cm}

\subsection{Task Description}
The UASpeech corpus \cite{kim2008dysarthric} is the largest publicly available dysarthric speech corpus consisting of 16 dysarthric and 13 control speakers, and contains 155 common words and 300 uncommon words. The entire corpus is further divided into 3 subset blocks per speaker. The same set of 155 common words is used in all three blocks, while the uncommon words in each block differ. The data from Block 1 and 3 of all the 29 speakers are used as the training set, while the data of Block 2 of all the 16 dysarthric speakers serves as the test set. After removing excessive silence and speed perturbation based data augmentation \cite{geng2020investigation}, a total of 130.1 hours of audio data is used as the training set, while 9 hours of speech is used for evaluation. \\
\noindent The DementiaBank Pitt \cite{becker1994natural} corpus is the largest publicly available elderly speech corpus designed for speech-based Alzheimer’s Disease (AD) diagnosis. It
contains about 33 hours of audio recorded over AD assessment interviews between 292 elderly participants and clinical investigators.
% It is further split into a 27.2-hour training set, a 4.8-hour development set and a 1.1-hour evaluation set for ASR system development. 
After silence stripping and data augmentation, the training data is increased to 58.9 hours, while the development and evaluation sets contain 2.5 hours and 0.6 hours of audio respectively. \textbf{There
is no overlapping between the elderly speakers in the training, development and evaluation sets.} The test data word coverage rates of UASpeech and DementiaBank are 61\% and 98.7\%.

% adaptive FT on training data
%% adapter arc. + adapter label (null, global, spk, seve., spk + seve.) + supervision (GT, ) 
% test-time
%% adapter arc. + adapter label (null, global, spk, seve., spk + seve.) + supervision (decoded from sys.1) 
% Fact.

\begin{table*}
    \setlength\tabcolsep{2pt}
    
    \centering
    \caption{Performance of Hubert and Wav2vec2-conformer systems constructed without or with structured speaker-deficiency adaptation on the test set of \textbf{UASpeech} dysarthric speech, as well as development (Dev.) and evaluation (Eval.) sets of \textbf{DementiaBank Pitt} elderly speech respectively. ``INV.'' and ``PAR.'' refer to clinical investigators and elderly participants.
%% ``Adapter Param.'' in short for Adaptor Parameters, where  
%% first number stands for the parameter size of speaker adapters and the second one is for the speech deficiency adapter.
Sys.2$^{*}$, 9$^{*}$ are the upper bound of Sys.2,9 respectively, which use the ground truth speech transcript as supervision during test time adaptation.
$\dag$, $^\ast$ and $^\triangle$ denote statistically significant (MAPSSWE \cite{gillick1989some},$\alpha$ = 0.05) differences obtained against the baseline fine-tuned, non-structured speaker or deficiency alone adapted ASR systems (Sys. 1,5,7).}
\vspace{-0.2cm}
\scalebox{0.7}{
\begin{tabular}{c|c|c|c|c|c|c|cc|cccc|c|c|cc|cc|c|c} 
\hline\hline
\multirow{3}{*}{Sys.} & \multicolumn{3}{c|}{Adaptive Training}                                      & \multicolumn{3}{c|}{Test Time Adapt.}                                                                 & \multicolumn{8}{c|}{UASpeech}                                                                                                     & \multicolumn{6}{c}{DementiaBank Pitt}                                                                                                                                         \\ 
\cline{2-7}\cline{8-21}
                      & \multirow{2}{*}{\begin{tabular}[c]{@{}c@{}}Ada.\\Arc.\end{tabular}} & \multirow{2}{*}{\begin{tabular}[c]{@{}c@{}}Ada.\\Label\end{tabular}} & \multirow{2}{*}{Sup.} & \multirow{2}{*}{\begin{tabular}[c]{@{}c@{}}Ada.\\Arc.\end{tabular} } & \multirow{2}{*}{\begin{tabular}[c]{@{}c@{}}Ada.\\Label\end{tabular}} & \multirow{2}{*}{Sup.}                                      & \multicolumn{7}{c|}{WER (\%)}                         & \multirow{2}{*}{\begin{tabular}[c]{@{}c@{}}Ada. Param.\\(spk.+defi.)\end{tabular}} & \multicolumn{2}{c|}{Dev. WER (\%)} & \multicolumn{2}{c|}{Eval. WER (\%)} & \multirow{2}{*}{O.A.} & \multirow{2}{*}{\begin{tabular}[c]{@{}c@{}}Ada. Param.\\(spk.+defi.)\end{tabular}}  \\ 
\cline{8-14}\cline{16-19}
                      &                               &                                &                      &                               &                                &                                                            & unseen & seen  & VL    & L     & M     & H    & O.A.  &                                                                           & Par.  & Inv.                       & Par.  & Inv.                        &                       &                                                                            \\ 
\hline\hline
1                     & \xmark                         & \xmark                          & \multirow{10}{*}{GT}  & \xmark                         & \xmark                          & \xmark                                                & 50.06  & 13.30 & 59.47 & 33.62 & 22.22 & 6.34 & 27.71 & -                                                                         & 29.73 & 14.28                      & 21.29 & 15.32                       & 21.61                 & -                                                                          \\ 
\cline{1-3}\cline{5-21}
2                     & \multirow{2}{*}{\xmark}                         & \multirow{2}{*}{\xmark}                          &                      & \multirow{2}{*}{RAB}                           & \multirow{2}{*}{global}                         &  Sys.1's outputs                                               & 47.08  & 13.61 & 59.91 & 32.40 & 19.73 & 5.51 & 26.73 & \multirow{2}{*}{0.5M}                                                                        & 29.40 & 14.20                      & 21.18 & 14.32                       & 21.39                 & \multirow{2}{*}{0.5M}                                                                         \\
\cline{1-1}\cline{7-7}
2$^{*}$                     &                       &                         &                      &                            &                         &  GT (supervised)                                          & 39.46 &  10.10  & 52.00  & 24.72 & 14.37
 & 4.26   & 21.61  &                                                                          & 28.19 & 13.86                      & 19.82 & 13.76                       & 20.53                 &                                                                           \\
\cline{1-3}\cline{5-21}
3                     & LHUC                          & spk.                           &                      & LHUC                          & spk.                           &    \multirow{7}{*}{{\begin{tabular}[c]{@{}c@{}} Sys.1's\\outputs\end{tabular}}}               & 47.91$^\dag$  & 13.31 & 60.02 & 31.36$^\dag$ & 20.98$^\dag$ & 5.96$^\dag$ & 26.88$^\dag$ & 0.016M                                                               & 29.01$^\dag$     & 14.39     & 20.93      & 13.76      & 21.25                & 0.147M                                                                 \\ 
\cline{1-3}\cline{5-6}\cline{8-21}
4                     & \xmark                         & \xmark                          &                      & \multirow{2}{*}{RAB}          & \multirow{2}{*}{spk.}          &                                     & 47.95$^\dag$  & 13.32 & 59.89 & 32.37 & 20.31$^\dag$ & 5.70$^\dag$ & 26.90$^\dag$ & \multirow{2}{*}{8M}                                                 &  28.83$^\dag$     &     14.02                       & 20.81      &   15.32                          &     21.06$^\dag$                  & \multirow{2}{*}{37M}                                                 \\
5                     & RAB                           & spk.                           &                      &                               &                                &                                                           & \cellcolor{orange!25}44.75$^\dag$  & \cellcolor{orange!25}12.46$^\dag$ & \cellcolor{orange!25}57.90$^\dag$  & \cellcolor{orange!25}29.02$^\dag$ & \cellcolor{orange!25}18.24$^\dag$ & \cellcolor{orange!25}5.46$^\dag$ & \cellcolor{orange!25}25.12$^\dag$ &                                                                           & \cellcolor{orange!25}28.06$^\dag$     & 13.84     & \cellcolor{orange!25}19.50$^\dag$       & 14.65      & \cellcolor{orange!25}20.45$^\dag$                &                                                                            \\ 
\cline{1-3}\cline{5-6}\cline{8-21}
6                     & \xmark                         & \xmark                          &                      & \multirow{2}{*}{RAB}          & \multirow{2}{*}{defi.}         &                                      &  47.93$^\dag$      &  13.10     &  59.59     &  32.17     &  20.65$^\dag$    &  5.45$^\dag$    &  26.76$^\dag$     & \multirow{2}{*}{2M}                                               &   29.03$^\dag$    &   14.36                         & 21.12      &       13.76                      &   21.28                    & \multirow{2}{*}{1M}                                                \\
7                     & RAB                           & defi.                          &                      &                               &                                &                                                            & \cellcolor{orange!25}45.41$^\dag$  & \cellcolor{orange!25}12.66$^\dag$ & 59.27 & \cellcolor{orange!25}29.79$^\dag$ & \cellcolor{orange!25}17.94$^\dag$ & \cellcolor{orange!25}5.30$^\dag$  & \cellcolor{orange!25}25.51$^\dag$ &                                                                            & \cellcolor{orange!25}29.13$^\dag$     & 14.50      & 20.13      & 14.43      & 21.23                &                                                                            \\ 
\cline{1-3}\cline{5-6}\cline{8-21}
8                     & \xmark                         & \xmark                          &                      & \multirow{3}{*}{\begin{tabular}[c]{@{}c@{}}Stru.\\RAB\end{tabular}}          & \multirow{3}{*}{\begin{tabular}[c]{@{}c@{}}spk.+\\defi.\end{tabular}}  &                                   & 47.08$^\dag$  & 13.21 & 59.32 & 32.16$^\dag$ & 19.67$^\dag$ & 5.38$^\dag$ & 26.49$^\dag$ & \multirow{3}{*}{\begin{tabular}[c]{@{}c@{}}10M\\(8M+2M)\end{tabular}}                                            &   28.52$^\dag$    &  14.29                          &   21.00    &   14.65                          &    21.05$^\dag$                   & \multirow{3}{*}{\begin{tabular}[c]{@{}c@{}}38M\\(37M+1M)\end{tabular}}                                            \\
9                     & Stru. RAB                           & spk.+defi.                   &                      &                               &                                &                                                           & \cellcolor{red!25}\textbf{44.60}$^{\dag\triangle}$          & \cellcolor{red!25}\textbf{11.87}$^{\dag\ast\triangle}$ & \cellcolor{red!25}\textbf{57.38}$^{\dag\triangle}$          & \cellcolor{red!25}\textbf{28.41}$^{\dag\ast\triangle}$                  & \cellcolor{red!25}\textbf{17.45}$^{\dag\ast}$      & \cellcolor{red!25}5.45$^\dag$          & \cellcolor{red!25}\textbf{24.70}$^{\dag\ast\triangle}$ &                                                                          & \cellcolor{red!25}\textbf{27.57}$^{\dag\triangle}$     & \cellcolor{red!25}\textbf{13.68}$^{\dag\triangle}$     & \cellcolor{red!25}\textbf{19.13}$^{\dag\triangle}$      & 14.32      & \cellcolor{red!25}\textbf{20.11}$^{\dag\triangle}$               &                                                                            \\ 
\cline{1-3}\cline{7-14}\cline{16-20}
9$^{*}$                    & Stru. RAB                           & spk.+defi.                  &                      &                               &                                &   GT (supervised)                                                        & 30.27        & 5.51  &  40.25       &    16.72             &  7.43    & 2.79          & 15.22  &                                                                          &  24.41   &  12.06    &  16.58    & 10.10      & 17.49              &                                                                            \\ 
%\cline{1-3}\cline{5-6}\cline{8-21} 
\hline\hline
% 0                     & \multicolumn{6}{c|}{3-way joint decoding TDNN system\cite{10584335}}                                                                                                                                                                 & 39.85  & 11.72 & 52.07 & 24.69 & 15.27 & 7.10 & 22.75 & -                                                                         & 26.05 & 12.52                      & 17.59 & 11.43                       & 18.69                 & -                                                                          \\ 
% \hline
1A                 & \multicolumn{6}{c|}{stronger SFM baseline \cite{10584335} using cross-system rescoring}                                                                                                                                               & 34.28  & 11.71 & 50.70 & 23.51 & 12.06 & 4.20 & 20.56 & -                                                                         & 25.27 & 12.07                      & 16.73 & 11.88                       & 18.07                 & -                                                                          \\
9A                & \multicolumn{6}{c|}{Sys. 1A + 9 (structured speaker-deficiency adaptation)}                                                                                                                                                                                                  & \cellcolor{blue!25}\textbf{33.17}  & \cellcolor{blue!25}\textbf{10.60}  & \cellcolor{blue!25}\textbf{49.34} & \cellcolor{blue!25}\textbf{21.60}  & \cellcolor{blue!25}\textbf{10.84} & \cellcolor{blue!25}\textbf{3.92} & \cellcolor{blue!25}\textbf{19.45}  & -                                                                        & \cellcolor{blue!25}\textbf{24.17}     & \cellcolor{blue!25}\textbf{11.88}     & \cellcolor{blue!25}\textbf{16.15}      & \cellcolor{blue!25}\textbf{11.54}      & \cellcolor{blue!25}\textbf{17.45}                  & -                                                                          \\
\hline\hline
\end{tabular}
}
\label{tab:main_res}
\vspace{-0.4cm}
\end{table*}

\subsection{Experimental Setup}
The pre-trained models on UASpeech and DementiaBank corpora are the large HuBERT model\footnote{huggingface.co/facebook/hubert-large-ls960-ft} and Wav2vec2-conformer model with relative position embeddings\footnote{huggingface.co/facebook/wav2vec2-conformer-rel-pos-large-960h-ft} respectively. \textbf{They are selected as the strongest baseline foundation model for each task, following the
prior work \cite{10584335}.}
The bottleneck dimensionality of the deficiency-conditioned RAB is empirically set as 256 for both UASpeech and DementiaBank. While the bottleneck dimensionality of the speaker identity-conditioned RAB is set to 256 for UASpeech and 128 for DementiaBank. This difference is due to UASpeech having more speaker-level data (2025 seconds per speaker) compared to DementiaBank (76 seconds per speaker). 

\vspace{-0.1cm}
\subsection{Implementation Issues \& Ablation Studies}
Two key implementation issues affecting the performance of both non-structured and structured SFM adaptation are investigated. These include: \textbf{a)} the position of the adapter; and \textbf{b)} the architecture of the adapter. From the ablation study results in Table \ref{tab:my_label}\footnote{The speech deficiency labels of the test set speakers are automatically predicted using the well-trained spectro-temporal embedding feature based
neural network classifiers \cite{geng23b_interspeech, geng2022speaker}, while test set speaker identity labels are provided during test-time unsupervised adaptation.}, several trends can be observed: \textbf{1)} Among all non-structured speaker-adapted or deficiency-adapted systems, inserting the LHUC and RAB adapters respectively after the CNN encoder and in the 2$^{nd}$ Transformer block produces the best performance (Sys.2 vs. 3-6; Sys.9 vs. 8,10; Sys. 13 vs. 12). \textbf{2)} The best-performing structured speaker-deficiency adapted system is obtained when both speaker and deficiency RAB adapters are located immediately after the CNN encoder (Sys.14 vs. 15-17)\footnote{This may be because the outputs of the lower positioned CNNs exhibit greater variability for structured adaptation to fully exploit, when compared with the outputs produced by the higher positioned Transformer blocks.}.

\vspace{-0.2cm}
\subsection{Main Results On UASpeech Dysarthric Data}
From the results in Table \ref{tab:main_res} on UASpeech dataset, there are several trends can be found: 
\textbf{1)} All non-structured speaker or deficiency alone adapted systems, and those using the proposed structured speaker-deficiency adaptation outperform the baseline in-domain multi-speaker data fine-tuned HuBERT and that using global (test set level) adapters (Sys.3-9 vs. 1,2). \\
% \textbf{2)} Among all non-structured speaker-adapted systems, inserting the LHUC and RAB adapters respectively after the CNN encoder and in the 2$^{nd}$ Transformer block produce the best performance (Sys.2 vs. 3-6; Sys.9 vs. 8,10). %\textcolor{red}{\sout{These two systems outperform the baseline fine-tuned HuBERT with statistically significant WER reductions of 2.59\% and 2.20\% absolute (9.35\% and 7.94\% relative) respectively (Sys.9, 13 vs. 1).}}
\textbf{2)} Non-structured RAB-based speaker or deficiency alone adapted systems (highlighted in orange)
outperform the baseline HuBERT with statistically significant WER reductions of 2.59\% and 2.20\% absolute (9.35\% and 7.94\% relative) respectively (Sys.5, 7 vs. 1). \\
\textbf{3)} The structured speaker-deficiency adapted system (highlighted in red) outperforms the baseline fine-tuned Large HuBERT model, as well as non-structured speaker or deficiency alone adapted systems by statistically significant WERs up to 3.01\% absolute (10.86\% relative, Sys.9 vs. 1,5,7). \\
\textbf{4)} Non-structured and structured SFM adaptive fine-tuning produce a more neutral and robust starting point than standard non-AFT models for test time unsupervised adaptation. It improves performance up to 1.79\% absolute (7.24\% relative, Sys.5,7,9 vs. 4,6,8). \\
%\textbf{5)} To fast adapt to a specific dysarthric speaker, the deficiency-dependent RAB obtained from the training data is kept, and only the speaker-dependent RAB is estimated. Such structured speaker-deficiency adapted system outperforms the unadapted HuBERT model, as well as the speaker-dependent system (Sys.18 vs. Sys.1,9).
% \textbf{6)} Considering the tradeoff between model complexity and performance, a structured LHUC-based speaker-adapted and RAB-based speech deficiency-adapted system is constructed, which outperforms the unadapted system, as well as corresponding speaker-adapted or speech deficiency-adapted systems (Sys.18 vs. Sys.1,2,13). 
\textbf{5)} After applying the proposed structured speaker-deficiency adaptation approach (Sys. 9A) on a stronger baseline SFM
\cite{hu2022exploring, 10584335} obtained using cross-system rescoring (Sys. 1A), the lowest published WER of 19.45\% on all the 16 dysarthric speakers (49.34\% on very low intelligibility, 33.17\% on unseen words, highlighted in blue) is obtained.
The performance improvements over the baseline in-domain data fine-tuned HuBERT (Sys. 9 vs 1) are also retained after cross-system rescoring based system combination (Sys. 9A vs. 1A).

Finally, the performance of our best system (Sys. 9A, Table \ref{tab:main_res}, in blue) is contrasted against recently published state-of-the-art results on UASpeech in Table \ref{table_2}. All the systems follow the same block-based training-evaluation protocol\footnote{Block 1+3 data used in training, all the 16 dysarthric speakers of Block 2 for evaluation \cite{christensen2013combining, sehgal2015model, xiong2019phonetic, kim2008dysarthric}.}.

\begin{table}[H]
\centering
\caption{WERs of published and our best system on \textbf{UASpeech}
% Recently published and our WER (\%) on \textbf{UASpeech}.
%\textcolor{red}{``15 spk'' stands for evaluation using only 15 speakers of the test set}, but the same 4 speakers are used in the ``very low'' (VL) intelligibility group.
}
\vspace{-0.2cm}
\scalebox{0.7}{
\begin{tabular}{c|c|c} 
\hline\hline
System                                                                   & VL      & All     \\ 
\hline\hline
Sheffield-2020 Fine-tuning CNN-TDNN speaker adaptation \cite{xiong2020source}                & 68.24 & 30.76  \\
CUHK-2021 NAS  DNN  +  Data  Aug.  +  LHUC-SAT  +  AV  fusion \cite{liu2021recent}        & 60.30 & 25.21  \\
CUHK-2022 DNN + Data Aug. + LHUC-SAT + AUV fusion \cite{hu2022exploiting}                    & 60.14 & 24.82  \\
CUHK-2022 DNN + Data Aug. + SBE Adapt + LHUC-SAT \cite{geng2022speaker}                             & 59.30 & 25.05  \\
% CUHK-2022 TDNN + spectral basic GAN + LHUC-SAT \cite{jin2022personalized}                        & 59.18 & 27.85  \\
% Nagoya-2022~WavLM + LSTM Decoder + CTC-A Loss               & 71.5  & 51.8   \\
BUT-2022 Wav2vec2.0 + fMLLR + xvectors \cite{baskar2022speaker}                                & 57.72 & 22.83  \\
Nagoya Univ.-2022 WavLM \cite{violeta2022investigating} & 71.50 & 51.80 \\
FAU-2022 Cross-lingual XLSR + Conformer \cite{hernandez22_interspeech} & 62.00 & 26.10 \\
CUHK-2023 Kaldi TDNN + VAE-GAN + LHUC-SAT \cite{jin2023adversarial} & 57.31 & 27.78 \\
CUHK-2024 HuBERT + sys. comb. \cite{10584335}  & 50.70 & 20.56  \\
\textbf{HuBERT + structured adapt. + sys. comb. (Sys. 9A, Table \ref{tab:main_res}, ours)} & \textbf{49.34} & \textbf{19.45} \\
\hline\hline
\end{tabular}
}
\label{table_2}
\vspace{-0.1cm}
\end{table}

\subsection{Results On DementiaBank Elderly Speech}
The following trends consistent with those found on the dysarthric UASpeech data in Table ~\ref{tab:main_res} are also observed on DementiaBank (\textbf{100\% test set speakers unseen in training}):
%% on the DementiaBank task:
\textbf{1)} Structured speaker-deficiency adaptation (highlighted in red) outperforms all of the comparable speaker or deficiency alone adapted systems (Sys.9 vs. 3,5,7), and the baseline fine-tuned Wav2vec2-conformer system by statistically significant WER reductions of 1.50\% absolute (6.94\% relative, Sys.9 vs. 1).
\textbf{2)} After applying the proposed structured speaker-deficiency adaptation approach (Sys. 9A) on a stronger baseline SFM \cite{hu2022exploring, 10584335} (Sys. 1A), a new state-of-the-art overall WER of 17.45\% is obtained on the combined DementiaBank development and evaluation sets (highlighted in blue).

% cognitive -> AD/CONTROL/INV.

\vspace{-0.1cm}
\section{Conclusion}
\vspace{-0.1cm}
This paper proposes novel structured speaker-deficiency adaptation approaches for SSL pre-trained SFMs on dysarthric and elderly speech data. Speaker and speech deficiency invariant SFMs were constructed in their supervised adaptive fine-tuning stage to reduce undue bias to training data
speakers, and serves as a more neutral and robust starting point for test time unsupervised adaptation.
Speech variability attributed to speaker identity and speech impairment severity, or aging induced neurocognitive decline, are modelled using separate adapters that can be combined together to model any seen or unseen speaker.
Consistent performance improvements are obtained over the baseline fine-tuned HuBERT and Wav2vec2-conformer models by statistically significant WER reductions of 3.01\% and 1.50\% absolute (10.86\% and 6.94\% relative) on the benchmark UASpeech and DementiaBank Pitt test sets respectively. New state-of-the-art WERs are also obtained on both tasks. 
Future research will study the rapid and online adaptation of pre-trained ASR models for dysarthric and elderly speakers.

\bibliographystyle{IEEEtran}
\bibliography{smallbib}

% Generated by IEEEtran.bst, version: 1.13 (2008/09/30)
\begin{thebibliography}{10}
\providecommand{\url}[1]{#1}
\csname url@samestyle\endcsname
\providecommand{\newblock}{\relax}
\providecommand{\bibinfo}[2]{#2}
\providecommand{\BIBentrySTDinterwordspacing}{\spaceskip=0pt\relax}
\providecommand{\BIBentryALTinterwordstretchfactor}{4}
\providecommand{\BIBentryALTinterwordspacing}{\spaceskip=\fontdimen2\font plus
\BIBentryALTinterwordstretchfactor\fontdimen3\font minus \fontdimen4\font\relax}
\providecommand{\BIBforeignlanguage}[2]{{%
\expandafter\ifx\csname l@#1\endcsname\relax
\typeout{** WARNING: IEEEtran.bst: No hyphenation pattern has been}%
\typeout{** loaded for the language `#1'. Using the pattern for}%
\typeout{** the default language instead.}%
\else
\language=\csname l@#1\endcsname
\fi
#2}}
\providecommand{\BIBdecl}{\relax}
\BIBdecl

\bibitem{sehgal2015model}
S.~Sehgal \emph{et~al.}, ``{Model adaptation and adaptive training for the recognition of dysarthric speech},'' in \emph{SLPAT}, 2015.

\bibitem{xiong2018deep}
F.~Xiong \emph{et~al.}, ``{Deep learning of articulatory-based representations and applications for improving dysarthric speech recognition},'' in \emph{ITG-Symposium}.\hskip 1em plus 0.5em minus 0.4em\relax VDE, 2018, pp. 1--5.

\bibitem{liu2021recent}
S.~Liu \emph{et~al.}, ``{Recent Progress in the CUHK Dysarthric Speech Recognition System},'' \emph{TASLP}, vol.~29, pp. 2267--2281, 2021.

\bibitem{ye2021development}
Z.~Ye \emph{et~al.}, ``{Development of the CUHK Elderly speech recognition system for neurocognitive disorder detection using the dementiabank corpus},'' in \emph{ICASSP}.\hskip 1em plus 0.5em minus 0.4em\relax IEEE, 2021, pp. 6433--6437.

\bibitem{geng2022speaker}
M.~Geng \emph{et~al.}, ``{Speaker adaptation using spectro-temporal deep features for dysarthric and elderly speech recognition},'' \emph{TASLP}, vol.~30, pp. 2597--2611, 2022.

\bibitem{yue2022acoustic}
Z.~Yue \emph{et~al.}, ``{Acoustic Modelling From Raw Source and Filter Components for Dysarthric Speech Recognition},'' \emph{TASLP}, vol.~30, pp. 2968--2980, 2022.

\bibitem{hu23b_interspeech}
S.~Hu \emph{et~al.}, ``{Exploiting Cross-Domain And Cross-Lingual Ultrasound Tongue Imaging Features For Elderly And Dysarthric Speech Recognition},'' in \emph{INTERSPEECH}, 2023, pp. 2313--2317.

\bibitem{hu2022exploiting}
S.~Hu, S.~Liu \emph{et~al.}, ``{Exploiting Cross Domain Acoustic-to-articulatory Inverted Features for Disordered Speech Recognition},'' in \emph{ICASSP}.\hskip 1em plus 0.5em minus 0.4em\relax IEEE, 2022, pp. 6747--6751.

\bibitem{konig2018fully}
A.~K{\"o}nig \emph{et~al.}, ``{Fully automatic speech-based analysis of the semantic verbal fluency task},'' \emph{Dementia and geriatric cognitive disorders}, 2018.

\bibitem{alzheimer20192019}
A.~Association, ``{2019 Alzheimer's disease facts and figures},'' \emph{Alzheimer's \& dementia}, 2019.

\bibitem{ferri2005global}
C.~P. Ferri \emph{et~al.}, ``{Global prevalence of dementia: a Delphi consensus study},'' \emph{The lancet}, vol. 366, pp. 2112--2117, 2005.

\bibitem{rudzicz2014speech}
F.~Rudzicz \emph{et~al.}, ``{Speech recognition in Alzheimer’s disease with personal assistive robots},'' in \emph{SLPAT}, 2014, pp. 20--28.

\bibitem{zhou2016speech}
L.~Zhou \emph{et~al.}, ``{Speech Recognition in Alzheimer's Disease and in its Assessment.}'' in \emph{INTERSPEECH}, 2016, pp. 1948--1952.

\bibitem{mirheidari2019dementia}
B.~Mirheidari \emph{et~al.}, ``{Dementia detection using automatic analysis of conversations},'' \emph{Computer Speech \& Language}, vol.~53, pp. 65--79, 2019.

\bibitem{10584335}
S.~Hu \emph{et~al.}, ``{Self-Supervised ASR Models and Features for Dysarthric and Elderly Speech Recognition},'' \emph{TASLP}, vol.~32, pp. 3561--3575, 2024.

\bibitem{kodrasi2020spectro}
I.~Kodrasi and H.~Bourlard, ``{Spectro-temporal sparsity characterization for dysarthric speech detection},'' \emph{TASLP}, vol.~28, pp. 1210--1222, 2020.

\bibitem{smith1987temporal}
B.~L. Smith \emph{et~al.}, ``{Temporal characteristics of the speech of normal elderly adults},'' \emph{JSLHR}, vol.~30, pp. 522--529, 1987.

\bibitem{baevski2020wav2vec}
A.~Baevski \emph{et~al.}, ``{wav2vec 2.0: A framework for self-supervised learning of speech representations},'' \emph{Advances in NeuralIPS}, vol.~33, pp. 12\,449--12\,460, 2020.

\bibitem{chen2022wavlm}
S.~Chen \emph{et~al.}, ``{Wavlm: Large-scale self-supervised pre-training for full stack speech processing},'' \emph{JSTSP}, vol.~16, pp. 1505--1518, 2022.

\bibitem{hsu2021hubert}
W.-N. Hsu \emph{et~al.}, ``{Hubert: Self-supervised speech representation learning by masked prediction of hidden units},'' \emph{TASLP}, vol.~29, pp. 3451--3460, 2021.

\bibitem{gandelsman2022testtime}
\BIBentryALTinterwordspacing
Y.~Gandelsman \emph{et~al.}, ``{Test-Time Training with Masked Autoencoders},'' in \emph{NIPS}, A.~H. Oh, A.~Agarwal, D.~Belgrave, and K.~Cho, Eds., 2022. [Online]. Available: \url{https://openreview.net/forum?id=SHMi1b7sjXk}
\BIBentrySTDinterwordspacing

\bibitem{sun2024learning}
Y.~Sun \emph{et~al.}, ``{Learning to (learn at test time): Rnns with expressive hidden states},'' \emph{arXiv preprint arXiv:2407.04620}, 2024.

\bibitem{anastasakos1996compact}
T.~Anastasakos \emph{et~al.}, ``{A compact model for speaker-adaptive training},'' in \emph{ICSLP}, vol.~2.\hskip 1em plus 0.5em minus 0.4em\relax IEEE, 1996, pp. 1137--1140.

\bibitem{gales1998maximum}
M.~J. Gales, ``{Maximum likelihood linear transformations for HMM-based speech recognition},'' \emph{Computer speech \& language}, vol.~12, no.~2, pp. 75--98, 1998.

\bibitem{swietojanski2016learning}
P.~Swietojanski \emph{et~al.}, ``{Learning hidden unit contributions for unsupervised acoustic model adaptation},'' \emph{TASLP}, vol.~24, pp. 1450--1463, 2016.

\bibitem{8682667}
X.~Xie \emph{et~al.}, ``{BLHUC: Bayesian Learning of Hidden Unit Contributions for Deep Neural Network Speaker Adaptation},'' in \emph{ICASSP}, 2019, pp. 5711--5715.

\bibitem{geng23_interspeech}
M.~Geng \emph{et~al.}, ``{On-the-Fly Feature Based Rapid Speaker Adaptation for Dysarthric and Elderly Speech Recognition},'' in \emph{INTERSPEECH}, 2023, pp. 1753--1757.

\bibitem{deng2023confidence}
J.~Deng \emph{et~al.}, ``{Confidence score based speaker adaptation of conformer speech recognition systems},'' \emph{TASLP}, vol.~31, pp. 1175--1190, 2023.

\bibitem{kim2008dysarthric}
H.~Kim \emph{et~al.}, ``{Dysarthric speech database for universal access research},'' in \emph{INTERSPEECH}, 2008, pp. 1741--1744.

\bibitem{becker1994natural}
J.~T. Becker \emph{et~al.}, ``{The natural history of Alzheimer's disease: description of study cohort and accuracy of diagnosis},'' \emph{Archives of neurology}, vol.~51, pp. 585--594, 1994.

\bibitem{hu2022exploring}
S.~Hu \emph{et~al.}, ``{Exploring Self-supervised Pre-trained ASR Models For Dysarthric and Elderly Speech Recognition},'' in \emph{ICASSP}.\hskip 1em plus 0.5em minus 0.4em\relax IEEE, 2023, pp. 1--5.

\bibitem{baskar2022speaker}
M.~K. Baskar \emph{et~al.}, ``{Speaker adaptation for Wav2vec2 based dysarthric ASR},'' in \emph{INTERSPEECH}, 2022, pp. 3403--3407.

\bibitem{9053725}
R.~Takashima, T.~Takiguchi, and Y.~Ariki, ``{Two-Step Acoustic Model Adaptation for Dysarthric Speech Recognition},'' in \emph{ICASSP}, 2020, pp. 6104--6108.

\bibitem{shor19_interspeech}
J.~Shor \emph{et~al.}, ``{Personalizing ASR for Dysarthric and Accented Speech with Limited Data},'' in \emph{INTERSPEECH}, 2019, pp. 784--788.

\bibitem{qi2023parameter}
J.~Qi and H.~V. hamme, ``{Parameter-efficient Dysarthric Speech Recognition Using Adapter Fusion and Householder Transformation},'' in \emph{INTRESPEECH}, 2023, pp. 151--155.

\bibitem{geng23b_interspeech}
M.~Geng \emph{et~al.}, ``{Use of Speech Impairment Severity for Dysarthric Speech Recognition},'' in \emph{INTERSPEECH}, 2023, pp. 2328--2332.

\bibitem{abdel2013fast}
O.~Abdel-Hamid and H.~Jiang, ``{Fast speaker adaptation of hybrid NN/HMM model for speech recognition based on discriminative learning of speaker code},'' in \emph{ICASSP}.\hskip 1em plus 0.5em minus 0.4em\relax IEEE, 2013, pp. 7942--7946.

\bibitem{thomas2022efficient}
B.~Thomas \emph{et~al.}, ``{Efficient adapter transfer of self-supervised speech models for automatic speech recognition},'' in \emph{ICASSP}.\hskip 1em plus 0.5em minus 0.4em\relax IEEE, 2022, pp. 7102--7106.

\bibitem{fan2022towards}
R.~Fan \emph{et~al.}, ``{Towards better domain adaptation for self-supervised models: A case study of child asr},'' \emph{JSTSP}, vol.~16, no.~6, pp. 1242--1252, 2022.

\bibitem{10023274}
Z.-C. Chen \emph{et~al.}, ``{Exploring Efficient-Tuning Methods in Self-Supervised Speech Models},'' in \emph{SLT}, 2023, pp. 1120--1127.

\bibitem{geng2020investigation}
M.~Geng \emph{et~al.}, ``{Investigation of Data Augmentation Techniques for Disordered Speech Recognition.}'' in \emph{INTERSPEECH}, 2020, pp. 696--700.

\bibitem{gillick1989some}
L.~Gillick and S.~J. Cox, ``{Some statistical issues in the comparison of speech recognition algorithms},'' in \emph{ICASSP}.\hskip 1em plus 0.5em minus 0.4em\relax IEEE, 1989, pp. 532--535.

\bibitem{christensen2013combining}
H.~Christensen \emph{et~al.}, ``{Combining in-domain and out-of-domain speech data for automatic recognition of disordered speech.}'' in \emph{INTERSPEECH}, 2013, pp. 3642--3645.

\bibitem{xiong2019phonetic}
F.~Xiong \emph{et~al.}, ``{Phonetic analysis of dysarthric speech tempo and applications to robust personalised dysarthric speech recognition},'' in \emph{ICASSP}.\hskip 1em plus 0.5em minus 0.4em\relax IEEE, 2019, pp. 5836--5840.

\bibitem{xiong2020source}
F.~Xiong, J.~Barker \emph{et~al.}, ``{Source domain data selection for improved transfer learning targeting dysarthric speech recognition},'' in \emph{ICASSP}.\hskip 1em plus 0.5em minus 0.4em\relax IEEE, 2020, pp. 7424--7428.

\bibitem{violeta2022investigating}
L.~P. Violeta, W.~C. Huang, and T.~Toda, ``{Investigating Self-supervised Pretraining Frameworks for Pathological Speech Recognition},'' in \emph{INTERSPEECH}, 2022, pp. 41--45.

\bibitem{hernandez22_interspeech}
A.~Hernandez \emph{et~al.}, ``{Cross-lingual Self-Supervised Speech Representations for Improved Dysarthric Speech Recognition},'' in \emph{INTERSPEECH}, 2022, pp. 51--55.

\bibitem{jin2023adversarial}
Z.~Jin \emph{et~al.}, ``{Adversarial data augmentation using VAE-GAN for disordered speech recognition},'' in \emph{ICASSP}, 2023, pp. 1--5.

\end{thebibliography}
\end{document}